# Mining Educational Data Using Classification to Decrease Dropout Rate of Students

Dr. Saurabh Pal

Department of Computer Applications, VBS Purvanchal University, Jaunpur – 222001 (U.P.), India

*Abstract*— In the last two decades, number of Higher Education Institutions (HEI) grows rapidly in India. Since most of the institutions are opened in private mode therefore, a cut throat competition rises among these institutions while attracting the student to got admission. This is the reason for institutions to focus on the strength of students not on the quality of education. This paper presents a data mining application to generate predictive models for engineering student's dropout management. Given new records of incoming students, the predictive model can produce short accurate prediction list identifying students who tend to need the support from the student dropout program most. The results show that the machine learning algorithm is able to establish effective predictive model from the existing student dropout data.

## I. INTRODUCTION

One of the biggest challenges that higher education faces is to improve student dropout rate. Student dropout is a challenging task in higher education [1] and it is reported that about one fourth of students dropped college after their first year [1-3]. Student dropout has become an indication of academic performance and enrolment management. Recent study results show that intervention programs can have significant effects on dropout, especially for the first year. To effectively utilize the limited support resources for the intervention programs, it is desirable to identify in advance students who tend to need the support most. In this paper, we describe the experiments and the results from a data mining techniques for the students of Institute of Engineering and Technology of VBS Purvanchal University, Jaunpur to assist the student dropout program on campus. In our study, we apply machine learning algorithm to analyse and extract information from existing student data to establish predictive model. The predictive model is then used to identify among new incoming first year students those who are most likely to benefit from the support of the student dropout program.

Data mining combines machine learning, statistics and visualization techniques to discover and extract knowledge. Educational Data Mining (EDM) carries out tasks such as prediction (classification, regression), clustering, relationship mining (association, correlation, sequential mining, and causal data mining), distillation of data for human judgment, and discovery with models [6]. Moreover, EDM can solve many problems based on educational domain. Data mining is non-trivial extraction of implicit, previously unknown and potentially useful information from large amounts of data. It is used to predict the future trends from the knowledge pattern.

The main objective of this paper is to use data mining methodologies to find students which are likely to drop out their first year of engineering. In this research, the classification task is used to evaluate previous year's student dropout data and as there are many approaches that are used for data classification, the Bayesian classification method is used here. Information like marks in High School, marks in Senior Secondary, students family position etc. were collected from the student's management system, to predict list of students who need special attention.

## II. BAYESIAN CLASSIFICATION

The Naïve Bayes Classifier technique is particularly suited when the dimensionality of the inputs is high. Despite its simplicity, Naive Bayes can often outperform more sophisticated classification methods. Naïve Bayes model identifies the characteristics of dropout students. It shows the probability of each input attribute for the predictable state.

A Naive Bayesian classifier [21] is a simple probabilistic classifier based on applying Bayesian theorem (from Bayesian statistics) with strong (naive) independence assumptions. By the use of Bayesian theorem we can write

$$P(Ci|X) = \frac{P(X|Ci)P(Ci)}{P(X)}$$

We preferred naive bayes implementation because:
- Simple and trained on whole (weighted) training data
- Over-fitting (small subsets of training data) protection
- Claim that boosting "never over-fits" could not be maintained.
- Complex resulting classifier can be determined reliably from limited amount of data

## III. NAIVE BAYESIAN CLASSIFICATION ALGORITHM

The naive Bayesian classifier works as follows:
- Let D be a training set of tuples and their associated class labels. As usual, each tuple is represented by an n-dimensional attribute vector, $X=(x_1, x_2,…, x_n)$, depicting n measurements made on the tuple from n attributes, respectively, $A_1, A_2,……, A_n$.
- Suppose that there are m classes, $C_1, C_2,…,C_m$. Given a tuple, X, the classifier will predict that X

belongs to the class having the highest posterior probability, conditioned on X. That is, the naïve Bayesian classifier predicts that tuple x belongs to the class Ci if and only if

$P(C_i|X) > P(C_j|X)$ for $1 \leq j \leq m, j \neq i$

Thus we maximize $P(C_i|X)$. The class $C_i$ for which $P(C_i|X)$ is maximized is called the maximum posteriori hypothesis. By Bayes' theorem

$$P(Ci|X) = \frac{P(X|Ci)P(Ci)}{P(X)}$$

- As P(X) is constant for all classes, only $P(X|C_i) P(C_i)$ need be maximized. If the class prior probabilities are not known, then it is commonly assumed that the classes are equally likely, that is, $P(C_1)=P(C_2)=\ldots= P(C_m)$, and we would therefore maximize $P(X|C_i)$. Otherwise, we maximize $P(X|C_i)P(C_i)$. Note that the class prior probabilities may be estimated by $P(C_i)=|C_i, D|/|D|$, where $|C_i, D|$ is the number of training tuples of class $C_i$ in D.

- Given data sets with many attributes, it would be extremely computationally expensive to compute $P(X|C_i)$. In order to reduce computation in evaluating $P(X|C_i)$, the naïve assumption of class conditional independence is made. This presumes that the values of the attributes are conditionally independent of one another, given the class label of the tuple (i.e., that there are no dependence relationships among the attributes). Thus,

$$P(X|C_i) = \prod_{k=1}^{n} P(X_k|C_i)$$

$=P(X_1|C_i)* P(X_2|C_i)*\ldots* P(X_m|C_i)$.

We can easily estimate the $P(X_1|C_i)* P(X_2|C_i)*\ldots* P(X_m|C_i)$ from the training tuples. Recall that here $X_k$ refers to the value of attribute $A_k$ for tuple X. For each attribute, we look at whether the attribute is categorical or continuous-valued. For instance, to compute $P(X|C_i)$, we consider the following:

- If $A_k$ is categorical, then $P(X_k|C_i)$ is the number of tuples of class $C_i$ in D having the value $x_k$ for $A_k$, divided by $|C_i, D|$, the number of tuples of class $C_i$ in D.

- If $A_k$ is continuous valued, then we need to do a bit more work, but the calculation is pretty straightforward. A continuous-valued attribute is typically assumed to have a Gaussian distribution with a mean μ and standard deviation σ , defined by

$$g(x, \mu, \sigma) = \frac{1}{\sqrt{2\pi}\sigma} e^{-\frac{(x-\mu)^2}{2\sigma^2}}$$

So that

$P(x_k|C_i)=g(x_k, \mu c_i, \sigma c_i)$

We need to compute $\mu c_i$ and $\sigma c_i$, which are the mean and standard deviation, of the values of attribute $A_k$ for training tuples of class $C_i$. We then plug these two quantities into the above equation.

- In order to predict the class label of X, $P(X|C_i)P(C_i)$ is evaluated for each class $C_i$. The classifier predicts that the class label of tuple X is the class $C_i$ if and only if

$P(X|C_i)P(C_i) > P(X|C_j)P(C_j)$ for $1 \leq j \leq m, j \neq i$

In other words, the predicted class label is the class $C_i$ for which $P(X|C_i)P(C_i)$ is the maximum.

## IV. BACKGROUND AND RELATED WORK

Tinto [2] developed the most popular model of retention studies. According to Tinto's Model, withdrawal process depends on how students interact with the social and academic environment of the institution.

To understand the factors influencing university student retention, Superby et. al. [9] used questionnaires to collect data including personal history of the student, implication of student behaviour and perceptions of the student. The authors applied different approaches such as decision tree, random forests, neural networks, and linear discriminate analysis to their questionnaires. However, possibly because of the small sample size, the prediction accuracy is not very good.

A number of Open Distance Learning institutions have carried out dropout studies. Some notable studies have been undertaken by the British Open University (Ashby [10]; Kennedy & Powell [11]). Different models have been used by these researchers to describe the factors found to influence student achievement, course completion rates, and withdrawal, along with the relationships between variable factors.

Yadav, Bharadwaj and Pal [12] conducted study on the student retention based by selecting 398 students from MCA course of VBS Purvanchal University, Jaunpur, India. By means of classification they show that student's graduation stream and grade in graduation play important role in retention.

Khan [13] conducted a performance study on 400 students comprising 200 boys and 200 girls selected from the senior secondary school of Aligarh Muslim University, Aligarh, India with a main objective to establish the prognostic value of different measures of cognition, personality and demographic variables for success at higher secondary level in science stream. The selection was based on cluster sampling technique in which the entire population of interest was divided into groups, or clusters, and a random sample of these clusters was selected for further analyses. It was found that girls with high socio-economic status had relatively higher academic achievement in science stream and boys with low socio-economic status had relatively higher academic achievement in general.

Pandey and Pal [14] conducted study on the student performance based by selecting 60 students from a degree college of Dr. R. M. L. Awadh University, Faizabad, India. By means of association rule they find the interestingness of student in opting class teaching language.

Ayesha, Mustafa, Sattar and Khan [15] describe the use of k-means clustering algorithm to predict student's learning activities. The information generated after the implementation of data mining technique may be helpful for instructor as well as for students.

Bharadwaj and Pal [16] obtained the university students data like attendance, class test, seminar and assignment marks from the students' previous database, to predict the performance at the end of the semester.

Bray [17], in his study on private tutoring and its implications, observed that the percentage of students receiving private tutoring in India was relatively higher than in Malaysia, Singapore, Japan, China and Sri Lanka. It was also observed that there was an enhancement of academic performance with the intensity of private tutoring and this variation of intensity of private tutoring depends on the collective factor namely socio-economic conditions.

Bhardwaj and Pal [18] conducted study on the student performance based by selecting 300 students from 5 different degree college conducting BCA (Bachelor of Computer Application) course of Dr. R. M. L. Awadh University, Faizabad, India. By means of Bayesian classification method on 17 attributes, it was found that the factors like students' grade in senior secondary exam, living location, medium of teaching, mother's qualification, students other habit, family annual income and student's family status were highly correlated with the student academic performance.

Yadav, Bharadwaj and Pal [19] obtained the university students data like attendance, class test, seminar and assignment marks from the students' database, to predict the performance at the end of the semester using three algorithms ID3, C4.5 and CART and shows that CART is the best algorithm for classification of data.

## IV. DATA MINING PROCESS

Knowing the reasons for dropout of student can help the teachers and administrators to take necessary actions so that the success percentage can be improved. Predicting the academic outcome of a student needs a lot of parameters to be considered. Prediction models that include all personal, social, psychological and other environmental variables are necessitated for the effective prediction of the performance of the students.

### A. Data Preparations

The data set used in this study was obtained from VBS Purvanchal University, Jaunpur (Uttar Pradesh) on the sampling method for Institute of Engineering and Technology for session 2010. Initially size of the data is 165.

### B. Data selection and Transformation

In this step only those fields were selected which were required for data mining. A few derived variables were selected. While some of the information for the variables was extracted from the database. All the predictor and response variables which were derived from the database are given in Table I for reference.

TABLE I
STUDENT RELATED VARIABLES

| Variables | Description | Possible Values |
|---|---|---|
| Branch | Students Branch | {CS, IT, ME} |
| Sex | Students Sex | {Male, Female} |
| Cat | Students category | {Unreserved, OBC, SC, ST} |
| HSG | Students grade in High School | {O – 90% -100%, A – 80% - 89%, B – 70% - 79%, C – 60% - 69%, D – 50% - 59%, E – 40% - 49%, F - < 40%} |
| SSG | Students grade in Senior Secondary | {O – 90% -100%, A – 80% - 89%, B – 70% - 79%, C – 60% - 69%, D – 50% - 59%, E – 40% - 49%, F - < 40% } |
| Atype | Admission Type | {UPSEE, Direct} |
| Med | Medium of Teaching | {Hindi, English} |
| LLoc | Living Location of Student | {Village, Town, Tahseel, District} |
| Hos | Student live in hostel or not | {Yes, No} |
| FSize | student's family size | {1, 2, 3, >3} |
| FStat | Students family status | {Joint, Individual} |
| FAIn | Family annual income status | {BPL, poor, medium, high} |
| FQual | Fathers qualification | {no-education, elementary, secondary, UG, PG, Ph.D., NA} |
| MQual | Mother's Qualification | {no-education, elementary, secondary, UG, PG, Ph.D., NA} |
| FOcc | Father's Occupation | {Service, Business, Agriculture, Retired, NA} |
| MOcc | Mother's Occupation | {House-wife (HW), Service, Retired, NA} |
| Dropout | Dropout: Continue to enroll or not after one year | {Yes, No} |

The domain values for some of the variables were defined for the present investigation as follows:

- Branch – The courses offered by VBS Purvanchal University, Jaunpur are Computer Science and Engineering (*CSE*), Information Technology (*IT*) and Mechanical Engineering (*ME*).

- Cat – From ancient time Indians are divided in many categories. These factors play a direct and indirect role in the daily lives including the education of young people. Admission process in India also includes different percentage of seats reserved for different categories. In terms of social status, the Indian population is grouped into four categories: Unreserved, Other Backward Class (OBC), Scheduled Castes (SC) and Scheduled Tribes (ST). Possible values are *Unreserved, OBC, SC and ST*.

- HSG - Students grade in High School education. Students who are in state board appear for six subjects each carry 100 marks. Grade are assigned to all students using following mapping *O – 90% to 100%, A – 80% - 89%, B – 70% - 79%, C – 60% - 69%, D – 50% - 59%, E – 40% - 49%, and F - < 40%.*

- SSG - Students grade in Senior Secondary education. Students who are in state board appear for five subjects

each carry 100 marks. Grade are assigned to all students using following mapping O – 90% to 100%, A – 80% - 89%, B – 70% - 79%, C – 60% - 69%, D – 50% - 59%, E – 40% - 49%, and F - < 40%.

- Atype - The admission type which may be through Uttar Pradesh State Entrance Examination (*UPSEE*) or *Direct* admission through University procedure.

- Med – This paper study covers only the colleges of Uttar Pradesh state of India. Here, medium of instructions are *Hindi or English*.

- FSize-. According to population statistics of India, the average number of children in a family is 3.1. Therefore, the possible range of values is from *1, 2, 3 or >3*.

- Dropout – Dropout condition. Whether the student continues or not after one year. Possible values are *Yes* if student continues study and *No* if student dropped the study after one year.

*C. Implementation of Mining Model*

Weka is open source software that implements a large collection of machine leaning algorithms and is widely used in data mining applications. From the above data, drop.arff file was created. This file was loaded into WEKA explorer. The classify panel enables the user to apply classification and regression algorithms to the resulting dataset, to estimate the accuracy of the resulting predictive model, and to visualize erroneous predictions, or the model itself. The algorithm used for classification is Naive Bayes. Under the "Test options", the 10-fold cross-validation is selected as our evaluation approach. Since there is no separate evaluation data set, this is necessary to get a reasonable idea of accuracy of the generated model. This predictive model provides way to predict whether a new student will continue to enroll or not after one year.

*D. Results and Discussion*

In the present study, those variables whose probability values were greater than 0.50 were given due considerations and the highly influencing variables with high probability values have been shown in Table 2. These values are calculated in the case of dropout student, who dropped the study after first year of engineering. These features were used for prediction model construction.

TABLE III
HIGH POTENTIAL VARIABLES

| Variable | Values | Probability |
|---|---|---|
| Sex | Male | 0.68 |
| SSG | E | 0.6623 |
| Atype | Direct | 0.6 |
| Med | Hindi | 0.76 |
| Lloc | Village | 0.55 |
| Mqual | Elementry | 0.50 |
| Mocc | Service | .52 |

Thable II indicates the most affective attributes were: Med, Sex, SSG, Atype, LLoc, Mocc and Mqual. Hints could be extracted from the table indicates that the students with Med = 'Hindi' are not continue their study. Wherever Sex is taken into consideration, Male students have greater possibility of discontinuation of study than Female.

The classification matrix has been presented in Table III, which compared the actual and predicted classifications. In addition, the classification accuracy for the two class outcome categories was presented.

TABLE III: CLASSIFICATION MATRIX-ID3 PREDICTION MODEL

| Dropout | | Predicted | |
|---|---|---|---|
| | | Yes | No |
| Actual | Yes | 121 | 10 |
| | No | 11 | 23 |

The class wise accuracy is shown in Table IV

TABLE IV: CLASS WISE ACCURACY

| Dropout | Precision | Recall |
|---|---|---|
| Yes | 0.917 | 0.924 |
| No | 0.697 | 0.676 |

V. CONCLUSIONS

Predicting students' dropout rate is great concern to the higher education system. Recently data mining can be used in a higher educational system to predict the list of students' who are going to drop their study. In this paper, we have introduced the Naïve Bayes classification algorithm for Student's dropout management in Higher Education system, to predict the relevancy of the incoming item from a new data set to the already existing data sets.

The empirical results show that we can produce short but accurate prediction of attributes for the student dropout purpose by applying the Naïve Bayes classification model to the records of incoming new students. This study will also work to identify those students which needed special attention to reduce drop-out rate.

INTERNATIONAL JOURNAL OF MULTIDISCIPLINARY SCIENCES AND ENGINEERING, VOL. 3, NO. 5, MAY 2012[7] Yoav Freund and Llew Mason, "The Alternating Decision Tree Algorithm". Proceedings of the 16th International Conference on Machine Learning, pages 124-133 (1999).

[8] Bernhard Pfahringer, Geoffrey Holmes and Richard Kirkby. "Optimizing the Induction of Alternating Decision Trees". Proceedings of the Fifth Pacific-Asia Conference on Advances in Knowledge Discovery and Data Mining. 2001, pp. 477-487.

[9] Superby, J.F., Vandamme, J-P., Meskens, N., "Determination of factors influencing the achievement of the first-year university students using data mining methods. Workshop on Educationa, 2006.

[10] Ashby, A., "Monitoring Student Retention in the Open University: Detritions, measurement, interpretation and action". *Open Learning, 19*(1), 65-78, 2004.

[11] Kennedy, D., & Powell, R., "Student progress and withdrawal in the Open University". Teaching at a Distance*, 7*, 61-78, 1976.

[12] Yadav S. K., Bharadwaj B. K. and Pal S., "Mining Educational Data to Predict Student's Retention: A Comparative Study", International Journal of Computer Science and Information Security (IJCSIS), Vol. 10, No. 2, Feb 2012, pp 113-117.

[13] Z. N. Khan, "Scholastic achievement of higher secondary students in science stream", Journal of Social Sciences, Vol. 1, No. 2, pp. 84-87, 2005.

[14] U. K. Pandey, and S. Pal, "A Data mining view on class room teaching language", (IJCSI) International Journal of Computer Science Issue, Vol. 8, Issue 2, pp. 277-282, ISSN:1694-0814, 2011.

[15] Shaeela Ayesha, Tasleem Mustafa, Ahsan Raza Sattar, M. Inayat Khan, "Data mining model for higher education system", Europen Journal of Scientific Research, Vol.43, No.1, pp.24-29, 2010.

[16] M. Bray, The shadow education system: private tutoring and its implications for planners, (2nd ed.), UNESCO, PARIS, France, 2007.

[17] B.K. Bharadwaj and S. Pal. "Mining Educational Data to Analyze Students' Performance", International Journal of Advance Computer Science and Applications (IJACSA), Vol. 2, No. 6, pp. 63-69, 2011.

[18] S. K. Yadav, B.K. Bharadwaj and S. Pal, "Data Mining Applications: A comparative study for Predicting Student's Performance", International Journal of Innovative Technology and Creative Engineering (IJITCE), Vol. 1, No. 12, pp. 13-19, 2011.
[ISSN: 2045-7057]                                                                      www.ijmse.org                                                                                           39